\documentclass[final]{ws-procs9x6}
\def\lb{\left(}
\def\rb{\right)}

\def\del{\delta}
\newcommand{\be}{\begin{equation}}
\newcommand{\ee}{\end{equation}}
\newcommand{\bea}{\begin{eqnarray}}
\newcommand{\eea}{\end{eqnarray}}

\begin{document}

\title{Dilaton Stabilization in Brane Gas
 Cosmology\footnote{\uppercase{T}his work is supported by the \uppercase{C}entre
    for \uppercase{H}igh \uppercase{E}nergy \uppercase{P}hysics at \uppercase{M}cGill, and \uppercase{NSERC}.}}

\author{A.~Berndsen\footnote{\uppercase{E}mail: aberndsen@physics.mcgill.ca}~ and~ J.~Cline\footnote{\uppercase{E}mail: jcline@physics.mcgill.ca}}
\address{Centre for High Energy Physics\\
 Ernest Rutherford Physics Building\\
McGill University\\
3600 rue University\\
Montréal, QC\\
Canada H3A 2T8}


\maketitle

\abstracts{
Brane Gas Cosmology is an M-theory motivated attempt to reconcile
aspects of the standard cosmology based on Einstein's theory of general
relativity. Dilaton gravity, when incorporating winding $p$-brane
states, has verified the Brandenberger--Vafa mechanism ---a
string-motivated conjecture which explains why only three of the nine
spatial dimensions predicted by string theory grow large. Further
investigation of this mechanism has argued for a hierarchy of
subspaces, and has shown the internal directions to be stable to initial
perturbations. These results, however, are dependent on a rolling
dilaton, or varying strength of Newton's gravitational constant
$G_N$. In these proceedings we show that it is not possible to
stabilize the dilaton and maintain the stability of the internal
directions within the standard Brane Gas Cosmology setup.}

\section{Introduction}
\subsection{Dilaton Gravity}\label{sec:dil}
Dilaton Gravity comes from the low-energy effective action of Type II-A
string theory, which is the result of M-theory compactified on $S^1$.  Ignoring contributions from the bulk anti-symmetric two-form and
including a potential $V(\phi)$ for the dilaton, the
action of this system is described as
\be
S=\int d^{10}x\sqrt{-G}e^{-2\phi}\lb
R+4G^{MN}\nabla_M\phi\nabla_N\phi-V(\phi)\rb\ ,\label{eq:iia}
\ee
where $G$ is the determinant of the background metric $G_{MN}$, $\phi$
is the dilaton, $R$ is the Ricci-scalar.  

Assuming a spatially-homogeneous dilaton and an FRW-type
metric of the
form~(\ref{eq:frw})\footnote{The space is described by two subspaces
  with the intention of one subspace describing the large directions ($\lambda$), and the
  other describing the remaining internal dimensions ($\nu$).}, the action reduces to
\bea
ds^2&=&-dt^2+e^{2\lambda(t)}\sum_{i=1}^3(dx_i)^2+e^{2\nu(t)}\sum_{j=1}^3(dy_j)^2\label{eq:frw}\\
S&=&\int d^{10}x\sqrt{-G_{00}}e^{-\varphi}G^{00}\lb
-3\dot\lambda^2-6\dot\nu^2+\dot\varphi^2-V(\phi)\rb\ ,
\eea
where the shifted dilaton $\varphi=2\phi-3\lambda-6\nu$ is introduced
for notational convenience.\cite{tv} Variation of the action with respect to
$G_{00}$, $\phi,\lambda$, and $\nu$ yields the system of equations
\bea
-3\dot\lambda^2-6\dot\nu^2+\dot\varphi^2&=&e^\varphi E + V(\phi)\label{eq:bgc1}\\
\ddot\varphi-3\dot\lambda^2-6\dot\nu^2&=&\frac{1}{2}e^\varphi
E+\frac{1}{4}V^\prime\nonumber\label{eq:bgc2}\\
\ddot\lambda-\dot\varphi\dot\lambda&=&\frac{1}{2}e^\varphi P_\lambda-\frac{1}{4}V^\prime\nonumber\label{eq:bgc3}\\
\ddot\nu-\dot\varphi\dot\nu&=&\frac{1}{2}e^\varphi
P_\nu-\frac{1}{4}V^\prime\label{eq:bgc4}\ .
\eea
The scale-factor equations of motion in~(\ref{eq:bgc4}) indicate one of the
substantial departures from general relativity: negative pressure
terms will cause deceleration\footnote{Recall that in general
 relativity negative pressure terms are normally associated with
 inflation, not deceleration.}.  Indeed, it is precisely this last
property that is exploited in Brane Gas Cosmology (BGC) to stabilize the internal
dimensions; BGC provides the negative pressure source\cite{bw}.
\par
In the standard approach to BGC\cite{abe,bek,bw}, energy $E$ and pressure $P_\lambda,\ P_\nu$ contributions come from $1$-branes, which are
described by the Dirac-Born-Infeld (DBI) action, while
$V(\phi)=0$.  The DBI action admits winding
and momentum states, whose energy and pressure contributions are
(respectively)
\bea
E_w=&\mu N e^\lambda,\hspace{2cm}P_w=&-\mu Ne^{\lambda}\nonumber\\ 
E_m=&\mu M e^{-\lambda},\hspace{2cm}P_m=&\mu Me^{-\lambda},
\eea
where $\mu$ is the brane tension, $N$ is the number of windings, and $M$
is the number of momentum modes.  The winding modes contribute the
negative pressure $P_w$ necessary to inhibit expansion of the internal
directions.

\subsection{The Rolling Dilaton}
We have remarked on one feature of dilaton-gravity which differs from
that of general relativity; if, however, this theory is to be consistent
with our current understanding of gravity, these differences must
disappear and we must recover general
relativity at some point in the early universe.
The transition occurs in the
case of a static dilaton: the action of dilaton-gravity, eq.~(\ref{eq:iia}), reduces to the
Einstein-Hilbert action, where the vacuum value of the dilaton
determines the strength of gravity as $G_N=e^{2\phi}/(16\pi)$.  Thus, a
rolling dilaton can be interpreted as a changing Newton's constant, which is
strongly constrained on experimental grounds.\cite{cfos}. 
 If BGC is
to make contact with our understanding of today's universe, the
dilaton must stabilize at some point of its evolution --- it is with this
motivation that we have introduced a dilaton potential into the
original action~(\ref{eq:iia}).

\section{Dilaton Stabilization}
\subsection{Perturbation Analysis}
To understand the effects of a stabilized dilaton we linearly perturb about a
static solution by $\phi=\phi_0+\del\phi,\
\lambda=\lambda_0+\del\lambda$, and $\nu=\nu_0+\del\nu$.  Such an
expansion should draw similar conclusions to a perturbation about a
slowly expanding universe since both the
radion and dilaton must have masses much greater than the eventual
Hubble rate.  Due to our
choice to expand about a static solution, and choosing
$\phi_0=\lambda_0=\nu_0=0$, the zeroth-order equations of the 
system~(\ref{eq:bgc4}) imply the relations
\bea
E&=&-V(\phi)\nonumber\\
E+3P_\lambda+6P_\nu&=&8V^\prime\nonumber\\
P_\lambda&=&\frac{1}{2}V^\prime\nonumber\\
P_\nu&=&\frac{1}{2}V^\prime\label{cc}
\eea
at the stationary point. 

To understand the effect of small fluctuations about the stationary point, we expand the
system~(\ref{eq:bgc4}) to first order.  Keeping in mind that
$E,\ P_\lambda$, and $P_\nu$ depend on $\lambda$ and $\nu$, we find
that 
\be
        \left(\begin{array}{c} \ddot\phi\\ \ddot\lambda\\ \ddot\nu
        \end{array} \right) = 
        {\bf S}
        \left(\begin{array}{c} \phi\\ \lambda\\  \nu\\
        \end{array} \right)
\ee
where the stability matrix $\bf S$ is given by 
\be
{\bf S} = 
\left(\begin{array}{ccc} 2(V'-\frac12V'')  & \ -\frac{25}{8} V'
        + \frac34 {\partial P_\lambda\over\partial\lambda} + 
        \frac32 {\partial P_\nu\over\partial\lambda}& \ -\frac{49}{8} V'
        + \frac34 {\partial P_\lambda\over\partial\nu} + 
        \frac32 {\partial P_\nu\over\partial\nu}\  \ \\
        \frac12 V'-\frac14V''& -\frac34 V' + \frac12{\partial
          P_\lambda\over\partial\lambda} 
        & -\frac32 V' + \frac12{\partial P_\lambda\over\partial\nu} \\
        \frac12 V'-\frac14V'' & -\frac34 V' + \frac12{\partial
          P_\nu\over\partial\lambda} 
        & -\frac32 V' + \frac12{\partial P_\nu\over\partial\nu}\\
                \end{array} \right) 
\ee

This matrix is diagonalized by a similarity transformation, $\bf P^{-1}SP$, so that
the general solution takes the form
\be
        \left(\begin{array}{c} \phi\\ \lambda\\  \nu\\
        \end{array} \right) = {\bf P} \left(\begin{array}{c} 
        A_1 e^{i\omega_1 t}\\ A_2  e^{i\omega_2 t}\\  A_3  e^{i\omega_3 t}\\
        \end{array} \right) + {(\omega_i\to -\omega_i)}
\ee
We insist that the eigenvalues $\omega_1$ and $\omega_3$ are real, giving a
stable dilaton and radion, while $\omega_2$ must be tuned to vanish (and the 
solution degenerates to $A_2 t + B$), so that the large dimensions are free to
expand when the tuning (\ref{cc}) is relaxed.  

However, the tuning of $\omega_2=0$ is not enough.  The matrix elements
$P_{\phi,2}$ and $P_{\nu,2}$ must be tuned to also vanish; otherwise $\phi$ and
$\nu$ will mix with the unstable mode and stability of the dilaton and radion
will be lost.  Certain linear combinations of $\phi$, $\nu$ and $\lambda$ will
oscillate, but this is not adequate: we really need $\phi$ and $\nu$ separately
to settle to some fixed values.  This can be accomplished by adjusting
parameters so that $S_{\phi,\lambda}=S_{\nu,\lambda}=0$.  Unfortunately, this
requires an exact cancellation between the first derivative of the dilaton
potential and quantities ($dP_\nu/d\lambda$ and $dP_\lambda/d\lambda$) characterizing the
brane gas, entities which have no reason to be related to each other.  If the
tuning fails by even a small amount, the contamination of the dilaton and
radion by the mode which expands like the scale factor of the large dimensions
will eventually lead to unacceptably large evolution of these fields. 

Let us contrast this to the usual BGC scenario where the dilaton is rolling and has no
potential.  In the radion-stability analysis\cite{bw}, the dangerous off-diagonal term
$S_{\nu,\lambda}=\frac12 dP_\nu/d\lambda$ is taken to be zero, and there is no
need to cancel it against $V'$.  The pressure in the extra dimensions is
assumed to be due to winding and momentum modes which exactly cancel each other
at $\nu=0$,
\be
        P_\nu = -\mu N (e^\nu - e^{-\nu}).
\ee
Here $\mu$ is the tension of the brane, and therefore positive,
and $N$ is the number of winding modes.
However $dP_\nu/d\nu$ is not zero, and has the right sign to stabilize the 
extra dimensions, since $S_{\nu,\nu} < 0$.  If we try to use the same
approach but also stabilize the dilaton, then $V=V'=P_\lambda=P_\nu=0$ at the stationary
point, while $V''$ can be nonzero.  We then have one vanishing eigenvalue,
as needed for the growth of the large dimensions, while the other two eigenvalues are given
by
\be
        -\omega^2_\pm = \frac12\left( \frac12 P' - V'' \pm
        \sqrt{\left(\frac12 P' - V''\right)^2 + \frac12 V''P'} \right)\label{eq:bgcev}
\ee
where $P'=dP_\nu/d\nu$. It can be shown that one of $\omega^2_\pm < 0$ for any value of
$P'$ or $V''$, meaning that there is always one unstable mode in
addition to the unstable $\omega=0$ mode. This result does not
contradict previous studies of the stability of the extra
dimensions in BGC because these works were only
concerned with the stability of the
radion\cite{bw,BWlinstab}. Instead, this result
shows that stabilization of the dilaton and the radion is
not possible without the introduction of some new potential for the
radion, and it demonstrates the importance of a rolling dilaton to the
stabilization mechanism in BGC.

\section{Conclusions}
The original proposal of Brandenberger and Vafa argues why, within a
string theory context, only three dimensions will be able to grow
large.\cite{bv} This idea has been extended to argue for a hierarchy of
subspaces, has been numerically verified, and issues such
the stability of the internal dimensions have been explored\cite{abe,bek,bw}. A
remaining aspect of this scenario is to understand the evolution of
the dilaton, an evolution which must stabilize at some later phase,
thus marking the transition to general relativity. To this end, we have introduced a potential for the dilaton
into the original action and tracked the evolution of perturbations
about a static configuration.
\par
The analysis indicates the importance of a rolling dilaton in order
for the negative-pressure winding modes to inhibit growth of the extra
dimensions. When the dilaton stabilizes, the original action reduces to
the usual Einstein-Hilbert action of general relativity, where all
sources act to accelerate expansion, causing the radion to grow. The eigenvalues of eq.~(\ref{eq:bgcev})
quantify this previous statement, indicating that any static
solutions are unstable to perturbations, so that two unstable modes
will always exist. Thus, solutions with one
growing mode (corresponding to three directions growing large), and
two stable modes (corresponding to a stable dilaton and radion) are
physically implausible within the normal BGC setup.
\par
Although BGC can stabilize the radion during the dilaton-gravity
epoch, some new mechanism must stabilize the internal dimensions once
the dilaton has stopped rolling.\cite{bc}  Presumably the onset of the
dilaton potential will coincide with the appearance of a potential for
the radion as well: one possibility is from compactifications with
fluxes\cite{gkp}.  This possibility provides an interesting link
between BGC and the work of Giddings, Kachru, and
Polchinski and subsequent investigations.


\section*{Acknowledgments}
We would like to thank Horace Stoica for helpful discussions.
A.~B. would like to acknowledge NSERC for their support in this
research via grant ES~B~267789, as well as support from the Physics Department at
McGill University.





\end{document}